\documentclass[%
 reprint,
 amsmath,amssymb,
 aps,
prb,
]{revtex4-2}

\usepackage{graphicx}
\usepackage{dcolumn}
\usepackage{bm}
\usepackage{hyperref}
\usepackage{braket}
\usepackage[bordercolor=white,backgroundcolor=gray!30,linecolor=black,colorinlistoftodos]{todonotes}
\usepackage{comment}

\hypersetup{
    unicode=false,          
    pdftoolbar=true,        
    pdfmenubar=true,        
    pdffitwindow=false,     
    pdfstartview={FitH},    
    pdftitle={Transverse-electric surface plasmon polaritons in periodically modulated graphene},    
    pdfauthor={Ahmad, Oh, Muljarov},     
    pdfsubject={},   
    pdfcreator={},   
    pdfproducer={}, 
    pdfkeywords={} {} {}, 
    pdfnewwindow=true,      
    colorlinks=true,       
    linkcolor=blue, 
    citecolor=blue,        
    filecolor=magenta,      
    urlcolor=blue,           
	breaklinks=true
}

\newcommand{\dis}{\displaystyle}

\newcommand{\Eq}[1]{Eq.\,(\ref{#1})}
\newcommand{\Eqs}[2]{Eqs.\,(\ref{#1}) and (\ref{#2})}

\newcommand{\Fig}[1]{Fig.\,\ref{#1}}
\newcommand{\Figs}[2]{Figs.\,\ref{#1} and \ref{#2}}
\newcommand{\Figsss}[3]{Figs.\,\ref{#1}, \ref{#2}, and \ref{#3}}
\newcommand{\Sec}[1]{Sec.\,\ref{#1}}

\newcommand{\be}{\begin{equation}}
\newcommand{\ee}{\end{equation}}
\newcommand{\bea}{\begin{eqnarray}}
\newcommand{\eea}{\end{eqnarray}}

\newcommand{\mat}[1]{\mathbf{#1}}
\newcommand{\I}{\mat{I}}

\newcommand{\W}{\mat{W}}
\newcommand{\Winv}{\mat{W}^{-1}}

\newcommand{\U}{\mat{U}}

\newcommand{\evalat}[2]{\left.#1\right|_{#2}}
\newcommand{\scatmat}{\mathbf{\mathbb{S}}}
\newcommand{\Vt}{\tilde{V}}

\begin{document}

\preprint{APS/123-QED}

\title{Transverse-electric surface plasmon polaritons in periodically modulated graphene}

\author{Zeeshan Ahmad}

\author{Sang Soon Oh}
\author{Egor A. Muljarov}\email{Egor.Muljarov@astro.cf.ac.uk}
\affiliation{School of Physics and Astronomy, Cardiff University, Cardiff CF24 3AA, United Kingdom}

\date{\today}
\begin{abstract}
Transverse-electric (TE) surface plasmon polaritons are unique eigenmodes of a homogeneous graphene layer that are tunable with the chemical potential and temperature. However, as their dispersion curve spectrally lies just below the light line, they  cannot be resonantly excited by an externally incident wave.
Here, we propose a way of exciting the TE modes and tuning their peaks in the transmission by introducing a one-dimensional graphene grating. Using the scattering-matrix formalism, we show that periodic modulations of graphene make the transmission more pronounced, potentially allowing for experimental observation of the TE modes. Furthermore, we propose the use of turbostratic graphene to enhance the role of the surface plasmon polaritons in optical spectra.
\end{abstract}

\maketitle

\section{\label{sec:level1}Introduction}

Surface plasmon polaritons (SPPs) in graphene have been an interesting area of theoretical and experimental research especially with the possibility of supporting SPPs with transverse-electric (TE) polarization in a graphene layer~\cite{Mikhailov2007}. More recently, TE SPPs were shown to exist in a homogeneous layer of graphene for an extended frequency range at non-zero temperature~\cite{Ahmad2021} using a complex frequency approach that models an open system with temporal decay. Although the TE SPP frequency dispersion is very close to the light line due to the small magnitude of the conductivity of graphene, proportional to the fine-structure constant~\cite{Mikhailov2007}, TE SPPs cannot be resonantly excited by an externally incident light because their dispersion curve lies below the light line.

Graphene is known for its tunability of optical conductivity by application of suitable gate voltage that induces sufficiently low but finite chemical potential~\cite{Lee2012}. This is because the electronic transitions occur near the K-point~\cite{Zhan2012} where the electronic dispersion is linear and the density of states vanishes. Devices like optical modulators~\cite{Shahram2020} and polarizers~\cite{Bao2011}, as well as absorption enhancement devices~\cite{Maji:18, guo2019} benefit from this tunability, which together with the existence of TE SPPs in graphene, provides exciting prospects for plasmonic applications~\cite{He2013}. In addition, structures with periodic graphene open the possibilities of generating topological plasmonic states~\cite{Kuzmin2018, Jin2017, Slipchenko2021, Taillefumier2011} when a magnetic field is applied.
Periodic plasmonic structures of graphene~\cite{gon2016,Cui2021, BenRhouma:22,Xiong2019}, and even multilayer stacks of periodic graphene strips~\cite{Emani2015, Li2019,Rodrigo2017,Yan2012,He2014} have been already studied. The effect of stacking graphene-dielectric layers on the properties of transverse-magnetic (TM) SPPs has also been analyzed for up to ten graphene layers~\cite{Gong2020}.
Although periodic graphene grating structure has been studied for TM SPP modes~\cite{gon2016,BenRhouma:22}, it has not been proposed for excitation and measurement of TE SPPs in graphene, to the best of our knowledge.

Various finite-element and finite-difference time-domain methods have been employed in many studies of plasmonic structures~\cite{Jin2017,HADDOUCHE2017132,MAIER2017126,Jung2007,shib2020,zang2020,rana2015,Olkkonen:10} to investigate SPPs. 
These fully numerical approaches are usually suitable for treating electromagnetic systems of relatively small sizes, but suffer from an increase in computation time for larger samples, since they require discretization of Maxwell's equations in real space over entire samples.
For larger samples and extended periodic systems, expanding fields into Fourier harmonics presents a more suitable approach~\cite{Weiss11}.  
One can approximate structural details by expanding the solutions to Maxwell's equations in the Fourier space~\cite{Antuu2011}, organizing the obtained solutions as coefficients of Bragg diffraction orders, which is a more appropriate approach for an infinite system with translational symmetries presented in this work. 
Although studies exist that consider periodic modulation of graphene at the atomic scale~\cite{Maksimova2012}, these are suitable for periods of the order of nanometers. For a modulation period of microns order, which is sufficiently large compared to the electronic wavelengths, we may approximate the optical conductivity of graphene as isotropic \cite{Nikitin2012} and consider the periodic modulation on the order of wavelengths of electromagnetic waves.

In this paper, we theoretically show that the TE SPP modes can be excited by incident light with an angle close to normal incidence with the help of one-dimensional periodic modulation in graphene. 
We demonstrate the excitation of the TE SPP modes as pronounced dips in the zeroth Bragg order transmission spectra. We also show that the in-plane wave number of the transmission dips are tunable with the grating period.
To enhance the transmission feature of the graphene grating, we propose the use of multilayer turbostratic graphene strips \cite{Wei2019,Baek2017}. 

To obtain the transmission spectra showing features due to the TE SPP modes, we employ the scattering-matrix formalism~\cite{Whittaker1999,TikhodeevPRB02} that is commonly used for systems periodically modulated in space. The fact that the width of a single- or multiple-layer graphene is much smaller than the SPP wavelength allows us to obtain an explicit analytical expression for the scattering matrix. The SPP modes, which are the eigen solutions of Maxwell's equations satisfying outgoing wave boundary conditions, correspond to the poles of the scattering matrix for the grating system~\cite{TikhodeevPRB02,Bykov2013} and manifest themselves as peaks and/or dips in transmission spectra. We have developed an analytical approximation for the scattering matrix, eigenmodes, and transmission, providing a proof that the features observed in optical spectra of periodically modulated graphene are in fact a manifestation of the TE SPP modes, which are unique to this kind of material.

\begin{figure}[]
\centering
\includegraphics[]{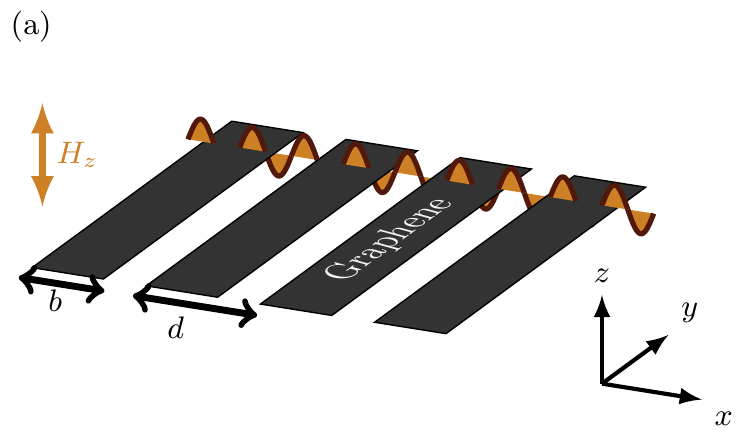}
\includegraphics[]{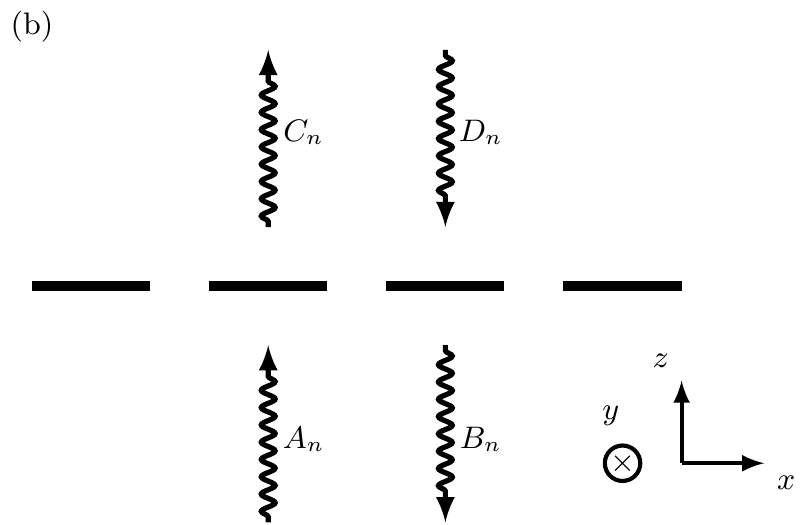}
\caption{Schematic of a graphene grating in a perspective view (a) and in the $xz$-plane side view (b). An infinitesimally thin graphene is periodically modulated along the $x$ axis at $z=0$ but homogeneous along the $y$ axis. Arrows labelled $A_n,B_n,C_n,D_n$ schematically represent incoming or outgoing coefficients of diffracted plane waves with an angle of incidence not necessarily normal.}
\label{fig:schematic}
\end{figure}

\section{Theoretical model}

In this section, we first derive, in the spirit of the scattering-matrix formalism~\cite{TikhodeevPRB02}, the transmission coefficients for a system containing a thin layer of a periodically modulated dispersive material surrounded by vacuum. Generalizations of the presented method to two-dimensional (2D) periodic modulation of a thin material layer and to systems with a substrate and/or superstrate are straightforward. The developed general formalism is then applied to a one-dimensional periodic array of single- and multiple-layer graphene strips, by using the available analytic expression for the conductivity of graphene. We also present here a useful approximation of the transmission at normal incidence, valid for the frequencies near the crossing point of the folded SPP mode dispersion, which allows us to analyse the optical response of the periodically modulated system in terms of its SPP modes.

In the following, we use the model of the permittivity which is described in the entire space as 
\begin{align}
    \varepsilon(x,z;\omega) = 1+\chi(\omega)\Lambda(x)\delta(z)\,,
    \label{struct_eps}
\end{align}
where $\chi(\omega)$ is the two-dimensional (2D) susceptibility of the infinitesimal layer,  $\Lambda(x)=\Lambda(x+d)$ is a periodic function with period $d$ describing the spatial modulation of the layer, and $\delta(z)$ is the Dirac delta function approximating the system as an infinitely thin layer. An illustration of the system, studied in this work, which is a periodic array of graphene strips, as well as the geometry of the electromagnetic field in the TE polarization are shown in \Fig{fig:schematic}(a). The use of the $\delta$ function is justified by the condition that the graphene thickness is much smaller than the wavelength of the electromagnetic field.

\subsection{Scattering matrix, transmission, and the eigenmodes of an inifnitely thin periodically modulated layer}
\label{sec-method}
The scattering matrix, transmission and reflection coefficients, as well as the secular equation determining the dispersion of SPP modes are obtained in this section by solving Maxwell's equations,
\begin{align}
    \nabla\times\mathbf{E}&=i\omega\mathbf{H}\,,\label{eqmaxwell1}\\
    \nabla\times\mathbf{H}&=-i\omega\varepsilon\mathbf{E}\,,\label{eqmaxwell2}
\end{align}
in which we assume non-magnetic materials and use the units in which the speed of light $c=1$. We also
assume a harmonic time dependence of the electromagnetic field in the form of $e^{-i\omega t}$ and use the frequency- and position-dependent permittivity $\varepsilon(x,z;\omega)$ introduced in \Eq{struct_eps}. For zero component of the light wave number in the $y$-direction, solutions of Maxwell's equations (\ref{eqmaxwell1}) and (\ref{eqmaxwell2}) split into TE and TM polarizations. We focus below on the TE polarization. The case of TM polarization for the same system is addressed in Appendix \ref{appendix_TM}.

For TE-polarized waves propagating in the $x$-direction [\Fig{fig:schematic}(a)] with the wave number $q$, the magnetic and electric fields in Cartesian coordinates take the form
\begin{align}
    \mathbf{H} = \begin{pmatrix}H_x\\0\\H_z\end{pmatrix}\,, \quad \mathbf{E} =\begin{pmatrix}0\\E_y\\0\end{pmatrix}\,,
\end{align}
respectively.
The periodic grating profile $\Lambda(x)$ can be written as a Fourier series,
\begin{align}
    \Lambda(x)=\sum_nV_n e^{ig_nx}\,,
\end{align}
where
\be
g_n=\frac{2\pi n}{d}
\label{eq_qn}
\ee
are the reciprocal lattice vectors and $V_n$ are Fourier coefficients of the periodic function $\Lambda(x)$. We can take the general electric-field solution to \Eqs{eqmaxwell1}{eqmaxwell2} in  the basis of the same Fourier harmonics,
\begin{align}
    E_y(x,z)=\sum_n e^{i(q+g_n)x}\times
    \begin{cases}
      C_n e^{ik_nz}+D_n e^{-ik_nz} & z> 0\,, \\
      A_n e^{ik_nz}+B_n e^{-ik_nz} & z< 0\,, \\
   \end{cases}
\label{eq_E}
\end{align}
where $A_n$, $B_n$, $C_n$, and $D_n$ are the field coefficients of the incoming and outgoing waves of the Bragg order $n$, as depicted in \Fig{fig:schematic}(b). The normal component of the light wave number for each Bragg order is given by
\be
k_n=\sqrt{\omega^2-(q+g_n)^2}\,,
\label{eq_kn}
\ee
with the square root in \Eq{eq_kn} chosen in such a way that either ${\rm Re}\,k_n>0$ or ${\rm Im}\,k_n>0$, for any real frequency $\omega$ and real wave number $q$, so the term $C_n e^{ik_nz}$ in \Eq{eq_E} describes, respectively, either a purely propagating in the $z$-direction or a purely evanescent wave. For complex frequencies, the choice of the square root is determined by a proper analytic continuation of \Eq{eq_kn} from the real $\omega$-axis into the complex frequency plane.

Integrating \Eqs{eqmaxwell1}{eqmaxwell2} across the infinitesimal grating layer results in an infinite set of simultaneous equations for the amplitudes  $A_n$, $B_n$, $C_n$, and $D_n$:
\begin{align}
     k_n(C_n\!-\!D_n\!-\!A_n\!+\!B_n)&=-\omega^2 \chi(\omega) \sum_{m}V_{n-m}(A_{m}\!+\!B_{m})\,,
     \nonumber\\
    C_n\!+\!D_n&=A_n\!+\!B_n\,,\label{eq_secular2}
\end{align}
see Appendix \ref{appendix_scat_mat} for a derivation.

Equations (\ref{eq_secular2}) can be written in matrix form as
\begin{align}
   \begin{pmatrix}
   \mathbf{C}\\\mathbf{B}
   \end{pmatrix} =\scatmat\begin{pmatrix}
   \mathbf{D}\\\mathbf{A}
   \end{pmatrix}\,, \label{eq_scatmat}
\end{align}
where $\scatmat$ is the scattering matrix, and $\mathbf{A}$, $\mathbf{B}$, $\mathbf{C}$, and $\mathbf{D}$ are vectors containing the field coefficients $A_n$, $B_n$, $C_n$, and $D_n$, respectively.
The scattering matrix can be expressed as
\begin{align}
    \scatmat=\begin{bmatrix} -\Winv\U& \Winv\\ \Winv&-\Winv\U\end{bmatrix} \,,\label{explicit_mat_result_main}
\end{align}
where $\Winv$ is the matrix inverse of $\W = \I+\U$, $\I$ is the identity matrix, and $\U$ is the matrix describing the coupling between diffraction orders, with the matrix elements given by
\begin{align}
    U_{nm}= \frac{\omega^2 \chi(\omega)}{2ik_n}V_{n-m}\,,
\end{align}
see Appendix \ref{appendix_scat_mat} for details.

For each Bragg channel $n$, the transmission coefficient $T_n$ is given by the ratio of the energy of the diffracted wave of order $n$ to the energy of an incoming wave (corresponding to $n=0$), which can be both evaluated from the normal component of the Poynting vector~\cite{TikhodeevPRB02}, which yields
\begin{align}
    T_n =\left|\frac{k_n}{k_0} \, C^2_n\right|\,.
    \label{transmission_exp}
\end{align}
Here, it is assumed without loss of generality that
\be
A_n= \delta_{n0}\,,\quad D_n=0\,,
\label{eq_BC}
\ee
where $\delta_{nm}$ is the Kronecker symbol, and the expansion coefficients $C_n$ in \Eq{transmission_exp} are found by solving the linear algebraic equations (\ref{eq_scatmat}) [or \Eq{eq_secular2}]  with the boundary conditions \Eq{eq_BC}. Note that the above definition of the transmission coefficients,  \Eq{transmission_exp},  is valid for so-called ``open'' Bragg channels only, for which
\begin{align}
    \omega^2>(q+g_n)^2 \,,\label{open_channel_condition}
\end{align}
equivalent to $k_n$ being real, see \Eq{eq_kn}. For ``closed" channels, the inequality \Eq{open_channel_condition} is not fulfilled, and the values of $T_n$ correspond to the near-field coefficients for these channels.

All the electromagnetic modes of the system, including the SPP modes and their dispersion relations, i.e. the dependence of the mode frequency on the wave number, can be found from \Eq{eq_scatmat} by applying the outgoing boundary conditions, $\mathbf{A}=\mathbf{D}=0$, for all diffraction orders. This is equivalent to solving
a secular equation
\begin{align}
    \text{Det[}\scatmat^{-1}(\omega;q)\text{]}=0\,, \label{eq_secular}
\end{align}
or finding the  poles of the scattering matrix $\scatmat(\omega;q)$ in the complex frequency plane for any real wave number $q$.

\begin{figure}[]
\centering
\includegraphics[]{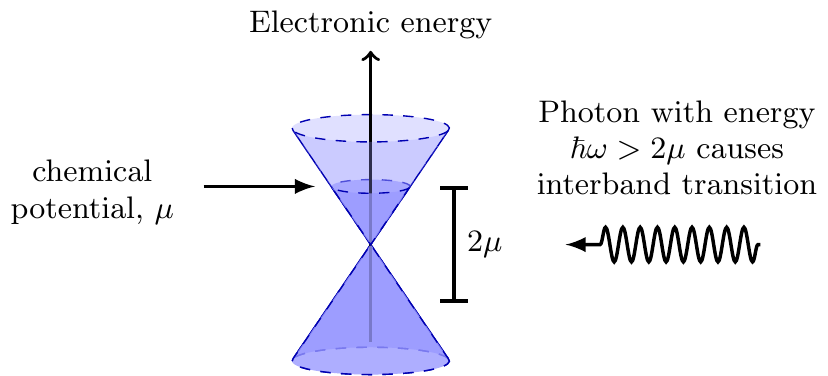}
\caption{Electronic dispersion of graphene near the K-point with non-zero chemical potential $\mu$ and zero temperature $T$.}
\label{dirac_cone}
\end{figure}

\subsection{Graphene conductivity and normalized frequencies and wave numbers}
While the general method introduced in \Sec{sec-method} is suitable for treating any thin periodically modulated layer, in what follows, we consider a graphene grating structure depicted in \Fig{fig:schematic}.
In this case, the susceptibility is given by
\begin{align}
    \chi(\omega) = \frac{2iN\sigma(\omega)}{\omega},
\label{eq_chi}
\end{align}
where $\sigma(\omega)$ is the 2D conductivity of a single homogeneous graphene layer and $N$ is the number of parallel graphene layers stuck together in such a way that, on the one hand, the total thickness of the system is much smaller than the SPP wavelength, but on the other hand, all the layers are electronically decoupled and periodically modulated with the modulation function $\Lambda(x)$, the same for all layers. For the graphene strips shown in \Fig{fig:schematic}, we take this periodic function in the following form
\begin{align}
 \Lambda(x)=\sum_n \Theta\left(\frac{b}{2}-|x-nd|\right)
\end{align}
where $\Theta(x)$ is the Heaviside step function, $d$ is the period, and $b$ is the width of the graphene strips.

\begin{figure}[]
\centering
\raisebox{12pt}{\includegraphics[]{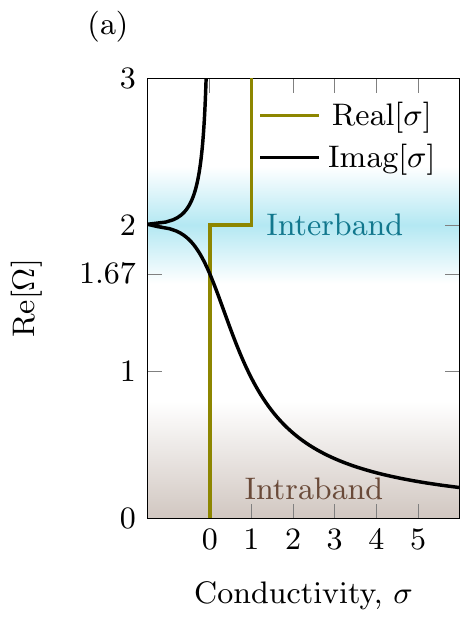}
}%
\includegraphics[]{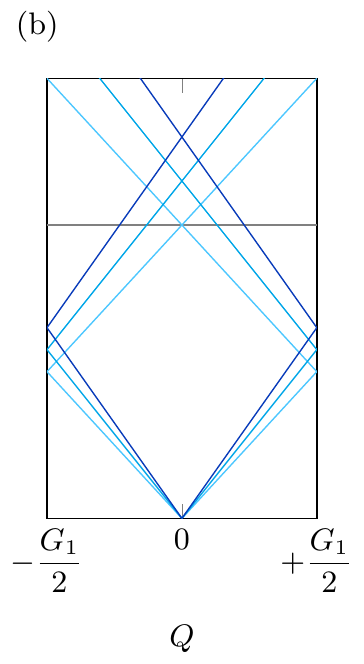}
\includegraphics[]{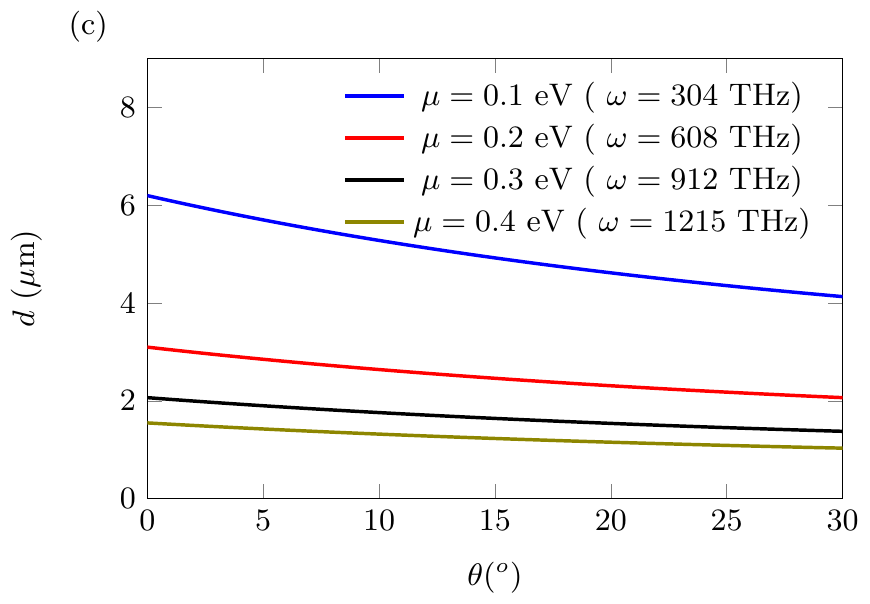}
\caption{(a) Conductivity of graphene at zero temperature $T=0$, sharing frequency axis with dispersion in panel (b). The frequency of interband pole in conductivity is affected by $\mu$ but not $d$. (b) Real part of the frequency of the TE mode dispersion of homogeneous graphene structure folded into the first Brillouin zone, shown for $G_1=2$, $2.3$ and $2.6$. The size of the first Brillouin zone, and consequently the frequency of mode crossing at $Q=0$, is controlled by $\mu d$. The TE mode frequency is close to the light line due to small value of the fine-structure constant. (c) The period of grating $d$ required for normalized frequency $\Omega=2$, various chemical potentials $\mu$ and corresponding frequencies $\omega$ (legend) as a function of desired incidence angle $\theta=\sin^{-1}(cq/\omega)$ in degrees, using \Eq{real_period_eq}.
}
\label{fig:folded_disp}
\end{figure}

In the long-wavelength limit, the single-layer frequency-dependent optical conductivity of graphene $\sigma(\omega)$ is presented in Refs.\,\cite{Mikhailov2007,Falkovsky_2008} and is shown in \Fig{fig:folded_disp}(a) for zero temperature. In terms of the notations used in Ref.\,\cite{Ahmad2021}, it is given by
\begin{align}
    \sigma(\omega)=2\pi[\sigma_\text{intra}(\omega)+\sigma_\text{inter}(\omega)]\,,
    \label{total_cond}
\end{align}
and full expressions for $\sigma_\text{intra}$ and $\sigma_\text{inter}$, the intraband and interband parts of the graphene conductivity, are derived in Ref.\,\cite{Ahmad2021}. While the interband part, given by the Drude model, describes the physical mechanism of conductivity  typical for normal metals, the interband part can be understood from the electronic band structure of graphene for a finite chemical potential $\mu$, as shown in \Fig{dirac_cone}. The interband part $\sigma_\text{inter}$ comes from optical transitions with energy greater than $2\mu$, contributing to a dip and a step-like feature at $\text{Re}\,\Omega=2$ which can be seen in \Fig{fig:folded_disp}(a), respectively, in the imaginary and real parts of the total conductivity \Eq{total_cond}.

Here and below we use for convenience dimensionless frequencies and wave numbers, all normalized to the chemical potential $\mu$ as
\be
\Omega=\frac{\hbar\omega}{\mu}\,,\quad Q= \frac{\hbar cq}{\mu}\,,\quad G_n= \frac{\hbar cg_n}{\mu}\,,\quad K_n= \frac{\hbar ck_n}{\mu}\,.
\ee
Equation (\ref{eq_kn}) then modifies to  $K_n=\sqrt{\Omega^2-(Q+G_n)^2}$, and similar changes are made in all other equations. The periodicity of the structure generates the first Brillouin zone with the range $Q\in[-G_1/2,G_1/2]$, as depicted in \Fig{fig:folded_disp}(b). The conductivity $\sigma(\omega)$ is then called below $\sigma(\Omega)$, which is a function of the normalized frequency $\Omega$, but it also depends  on a normalized inverse temperature $\mu\beta = \mu/k_BT$, where $k_B$ is the Boltzmann constant.

Let us note finally that multilayer graphene structures generally have interlayer coupling which alters the electronic band structure near the K-point and thus qualitatively changes the optical conductivity spectrum.
However, it is known that turbostratically stacked graphene layers have negligible interlayer coupling \cite{Min2008}, simply factorizing the graphene conductivity \cite{Baek2017}.
Such graphene stacks have already been synthesized \cite{Garlow2016,Wei2019}.
We propose to use turbostratic multilayer graphene strips \cite{Negishi2017} to enhance features in transmission due to the TE SPP mode, including the increase of its linewidth and the frequency gap in SPP mode dispersion. In this work, we obtain transmission spectra and approximate frequencies of the TE modes for both single-layer ($N=1$) and turbostratic ($N=10$) graphene grating by using the suitable number of graphene layers $N$ in \Eq{eq_chi}.

\subsection{Controlling mode frequency and wave number}

The conductivity spectrum of graphene has a key frequency $\text{Re}\,\Omega=2$ where the interband dip occurs at low temperatures, see \Fig{fig:folded_disp}(a). Since the scattering matrix depends on frequency $\omega=\mu\Omega/\hbar$
and the SPP mode coupling due to the periodic modulation of the system depends on the grating period $d$,
one  can control this coupling and the SPP mode properties by varying the chemical potential $\mu$ and the period $d$. In fact, by tuning the grating period $d$, the edge of the first Brillouin zone $G_1/2$ where the bands are folded also changes, leading to variations of the frequency of the TE SPP folded mode crossing at $Q=0$, as it is clear from \Fig{fig:folded_disp}(b).
Within and above the micrometer range, the grating period has a negligible effect on the conductivity spectrum, since the photon wave numbers in this case are much smaller than the electronic wave numbers~\cite{Ahmad2021} --- this regime of a negligible spatial dispersion is called above the long-wavelength limit of the conductivity. One can therefore independently choose a particular part of the conductivity spectrum shown in \Fig{fig:folded_disp}(a) and the desired normalized coupled mode frequency, where the two TE SPP dispersion lines (which are close to the light lines) cross. These are, for example, $\Omega=1.667$ and $\Omega=2$ in \Fig{fig:folded_disp}(a) and  $\Omega=2$, $2.3$, and $2.6$ in \Fig{fig:folded_disp}(b). For a better description of this tuning process, we define the normalized interband detuning $\Delta= G_1-2$, between the crossing in dispersion at $\Omega=G_1=2\pi\hbar c /(\mu d)$ and the interband dip in the imaginary part of conductivity at $\Omega=2$ as illustrated in \Fig{fig:folded_disp}(a).
As an example, for a chemical potential of $\mu$ = 0.2\,eV, the zero detuning $\Delta = 0$ would require the graphene stripe width to be in the order of micrometers. Since the dispersion of the TE mode is close to the light line~\cite{Mikhailov2007,Ahmad2021}, we may approximate the TE mode dispersion as the folded light line and estimate the grating period required to tune the TE mode to a desired frequency $\Omega$, for any in-plane wave number $Q$.
In \Fig{fig:folded_disp}(c) we show the dependence of the grating period $d$ required for the TE SPP mode of the $n=-1$ diffraction order to reach the frequency $\Omega=2$. For general $\Omega$ and $Q$ this condition is given by
\begin{align}
    d=\frac{\hbar c}{\mu} \,\frac{2\pi}{\Omega+Q}\,.
    \label{real_period_eq}
\end{align}
Note that this dependence is shown in \Fig{fig:folded_disp}(c) in terms of different values of the chemical potential $\mu$ in eV, unnormalized frequency $\omega$ in Hz, and the angle of incidence $\theta$ measured from the normal to the graphene plane and defined as $\sin\theta=Q/\Omega$. Clearly $|\theta|\leqslant 30^\circ$ in this case, since $\Omega= G_1$ and $|Q|\leqslant G_1/2$ within the first Brillouin zone.

\subsection{Role of the TE SPP modes in the transmission: Analytic approximation at normal incidence}
\label{sec:approximation}

To reveal the effects of periodic modulation on TE SPP modes and the role of these modes in the transmission, we focus in this section on the normal incidence and develop approximate analytical expressions for the transmission around the frequency Re\,$\Omega= G_1$. To do so, we truncate $\W$ in the scattering matrix [\Eq{explicit_mat_result_main}] to the lowest three diffraction orders, $n=0$ and $n=\pm1$:
\begin{align}\label{truncscat}
  \W = \begin{pmatrix}\dis
        1+\frac{\Omega\Vt_0}{K_{-1}} &\dis \frac{\Omega\Vt_{-1}}{K_{-1}} & \dis\frac{\Omega\Vt_{-2}}{K_{-1}}\\
        \dis \frac{\Omega\Vt_{1}}{K_{0}} & \dis 1+ \frac{\Omega\Vt_{0}}{K_{0}} & \dis \frac{\Omega\Vt_{-1}}{K_{0}}\\
        \dis \frac{\Omega\Vt_{2}}{K_{1}} & \dis \frac{\Omega\Vt_{1}}{K_{1}} & \dis 1+\frac{\Omega\Vt_{0}}{K_{1}}
    \end{pmatrix},
\end{align}
where $\Vt_n= N\sigma(\Omega)V_n$ and $K_n\equiv\sqrt{\Omega^2-(Q+G_n)^2}$.

The transmission coefficients are then given by the middle column entries of $\Winv$,
\begin{align}
    T_{\pm 1}=\frac{F_{\pm 1}}{\left|\W\right|}\,,\quad T_{0}=\frac{F_{0}}{\left|\W\right|}\,,\label{tcoeffthreemat}
\end{align}
where
\begin{align}
    F_{\pm 1}&=\frac{\Omega^2}{K_1K_{-1}}\Vt_{\mp 1}\Vt_{\pm 2}-\left(1+\frac{\Omega}{K_{\mp 1}}\Vt_0\right)\frac{\Omega}{K_{\pm 1}}\Vt_{\pm 1}\,,
\end{align}
\begin{align}
    F_{0}&=\left(1+\frac{\Omega}{K_{-1}}\Vt_0\right)\left(1+\frac{\Omega}{K_{1}}\Vt_0\right)-\frac{\Omega^2}{K_1K_{-1}}\Vt_2\Vt_{-2}\,,
\end{align}
and the determinant of \Eq{truncscat} is given by
\be
    \left|\W\right| = \left(1+\frac{\Omega}{K_0}\Vt_0\right)F_0+\frac{\Omega}{K_0}\Vt_1F_{-1}+\frac{\Omega}{K_0}\Vt_{-1}F_1\,.
\label{Wdet}
\ee
Note that the modulated TE SPP mode frequencies correspond to the zeros of this determinant.

For $Q=0$, corresponding to normal incidence, the approximate transmission coefficients obtained in \Eq{tcoeffthreemat}, can be further simplified by using the fact that $K_0=\Omega$ and $K_{\pm1}=\sqrt{\Omega^2-G_1^2}$, that results in
\bea
    F_{\pm1}&=&-\frac{\Omega \Vt_1}{K_1}\, \mathcal{D}_-(\Omega)\,,\\
    F_0&=&\mathcal{D}_-(\Omega)\,{\mathcal{D}_+}(\Omega)\,,\\
    \left|\W\right|&=&\mathcal{D}_-(\Omega)\,\mathcal{B}(\Omega)\,,
    \label{detWatzeroq}
\eea
where
\be
    \mathcal{D}_{\pm}(\Omega)=1+\frac{\Omega}{K_1}\left(\Vt_0\pm\Vt_2\right)\,,
\ee
 and
\be
    \mathcal{B}(\Omega)=1+\Vt_0+\frac{\Omega}{K_1}\left[\Vt_0+\Vt_2+\Vt_0\left(\Vt_0+\Vt_2\right)-2\Vt_1^2\right]\,.
\ee

Here,  without loss of generality, we have made use of the fact that the Fourier coefficients $V_n$ of the grating profile function $\Lambda(x)$ are real, if $\Lambda(x)$ is an even function of $x$; therefore $\Vt_n=\Vt_{-n}$. Thus, for $Q=0$, the transmission coefficients in \Eq{tcoeffthreemat} reduce to
\be
    T_0(\Omega)=\frac{{\mathcal{D}_+}(\Omega)}{\mathcal{B}(\Omega)}\,,\quad
    T_{\pm1}(\Omega)=-\frac{\Omega}{K_1}\frac{\Vt_1}{\mathcal{B}(\Omega)}\,.
\label{approxt0}
\ee

As it is clear from \Fig{fig:folded_disp}(b), the two folded dispersion lines originating from the unperturbed TE SPP of the homogeneous layer, corresponding to the $n=1$ and $n=-1$ Bragg channels and crossing at $\Omega\approx G_1$, produce in a periodically modulated system two perturbed TE SPP modes at each $Q$. The frequencies of these modes can be found from the zeros of the determinant \Eq{Wdet}, which factorizes at $Q=0$,  according to \Eq{detWatzeroq}. One mode, having frequency $\Omega_d$, is found from the condition $\mathcal{D}_-(\Omega_d)=0$, which yields
\be
\frac{\Omega_d}{G_1}=\frac{1}{\sqrt{\dis 1-\left(\Vt_0-\Vt_2\right)^2}}\,,
\label{darkzero}
\ee
and is further referred to as {\it dark mode}, as it does not manifest itself in the normal-incidence transmission. In fact, since $F_{\pm1}$ and $F_0$ are proportional to $\mathcal{D}_-(\Omega)$ there is an exact cancellation of this factor in $T_{\pm1}$ and $T_0$. The other mode, having frequency $\Omega_b$, which we further call {\it bright mode}, is obtained from equating to zero the other factor in $|\W|$ given by \Eq{detWatzeroq}, $\mathcal{B}(\Omega_b)=0$, which yields
\be
    \frac{\Omega_b}{G_1}=\frac{1+\Vt_0}{\sqrt{\dis\left(1+\Vt_0\right)^2-\left[(1+\Vt_0)(\Vt_0+\Vt_2)-\Vt_1^2\right]^2}}
    \label{brightzero}\,,
\ee
This mode does contribute to the transmission, which in turn can be approximated for the frequencies $\Omega$ close to $\Omega_b$ as
\be
    T_0(\Omega)\approx
    \frac{\mathcal{D}_-(\Omega_b)+(\Omega-\Omega_b)\mathcal{D}_-'(\Omega_b)}{(\Omega-\Omega_b)\mathcal{B}'(\Omega_b)}
    \,,\label{tnotapprox}
\ee
and a similar approximate expression can be obtained also for $T_{\pm1}(\Omega)$.

\begin{figure*}[t]
\centering
\includegraphics[]{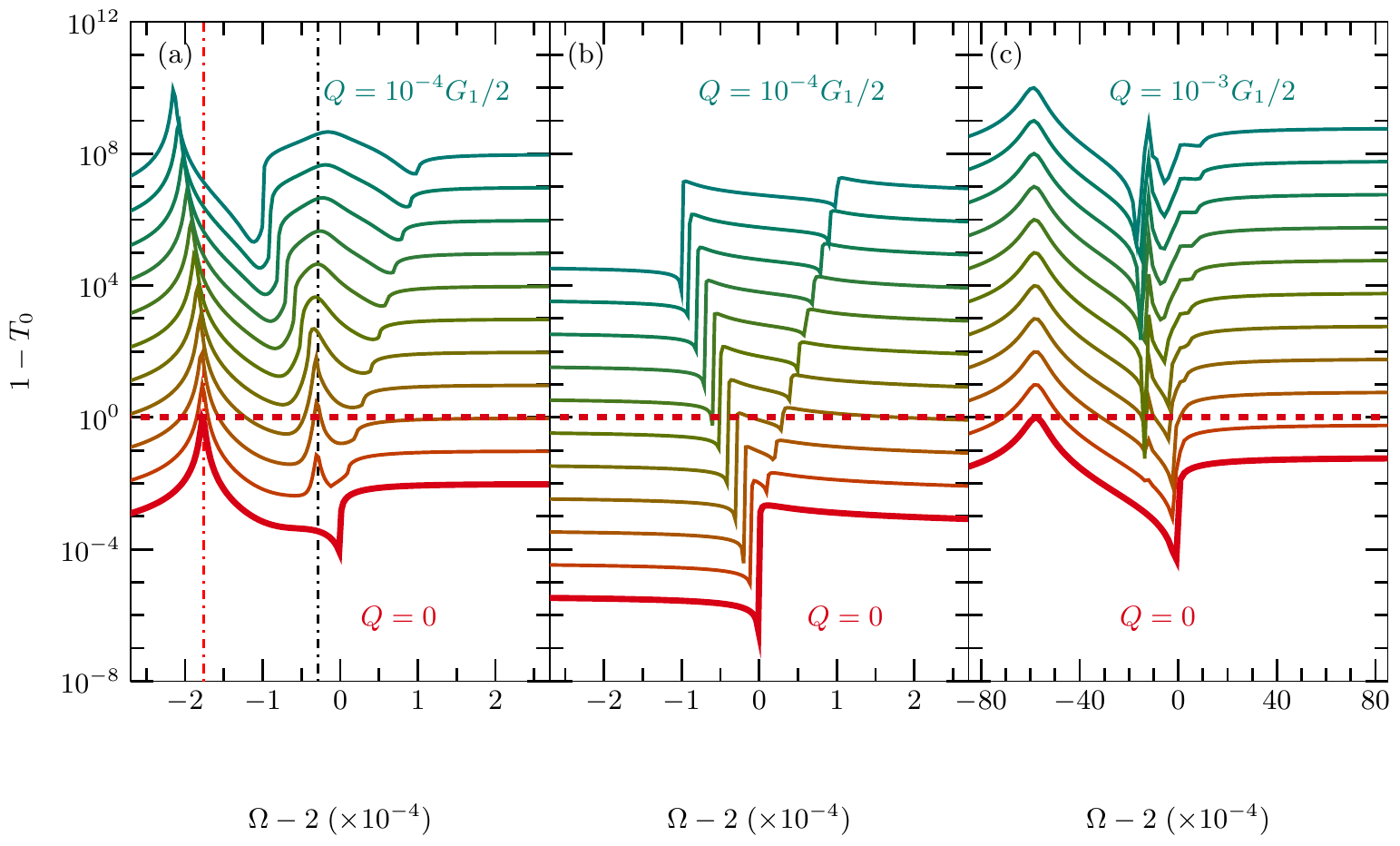}
\caption{Transmission spectra ($1-T_0$) for zeroth diffraction order (close-to-normal incidence) when a TE polarised light with parallel wave number $Q$, is incident on the grating with $b/d=0.3$, $\Delta=0$, for (a) single-layer graphene with $\sigma/2\pi=\sigma_\text{intra}+\sigma_\text{inter}$, (b) single-layer graphene without interband conductivity  $\sigma/2\pi=\sigma_\text{intra}$, and (c) turbostratically stacked graphene with $\sigma/2\pi= N(\sigma_\text{intra}+\sigma_\text{inter})$, for $N=10$. The dashed red line indicates $T_0=0$. Spectra for $Q\neq 0$ are offset by powers of $\times 10$ in increasing order of $Q$. Vertical dash-dotted lines are at TE SPP resonance frequencies for $Q=0$.}
\label{fig:enhance}
\end{figure*}

Note that
\Eqs{darkzero}{brightzero} are not explicit expressions for, respectively, $\Omega_d$ and $\Omega_b$, but should instead be solved self-consistently, as the matrix elements $\Vt_n$, even though they are small, contain dependence on frequency, which is present in the conductivity $\sigma(\Omega)$. However, one can use $\Omega=G_1$ as a good starting value of the argument of $\Vt_n(\Omega)$ for a quickly converging self-consistent iterative solution of \Eqs{darkzero}{brightzero}, which is implemented in this work and illustrated below in \Sec{resultssec}.

Note also that in the absence of the periodic modulation, $\Vt_n=\delta_{n,0}N\sigma(\Omega)$, and both modes have exactly the same frequency, given by 
$\Omega_{b,d}= {G_1}/{\sqrt{1-\Vt_0^2}}$, which 
coincides with the homogeneous grahene dispersion~\cite{Ahmad2021} folded into the center of the first Brillouin zone.

\section{Results} \label{resultssec}

We now present transmission spectra of periodically modulated graphene at $\mu\beta=10^5$, with $30\%$ filling factor ($b/d=0.3$), for normal and close-to-normal incidence of a TE-polarized incoming plane wave. Here, the grating period $d$ is chosen in such a way that $\Delta=0$ (equivalent to $G_1=2$). For this grating period, the crossing point of the folded light dispersion at $Q=0$ coincides  with the interband dip in the imaginary part of the conductivity, see Figs.\,\ref{fig:folded_disp}(a) and (b). It is therefore expected that the coupling of the  TE SPP modes of a homogeneous graphene, which is caused by its periodic modulation, is maximized, and thus features in transmission which are due to the TE SPP modes are enhanced.

\Fig{fig:enhance} shows the transmission spectra in the zeroth diffraction order, $1-T_0$,  for single-layer graphene grating in panel (a), single-layer graphene grating without interband conductivity $\sigma_\text{inter}$ in panel (b), and turbostratic ($N=10$) graphene grating in panel (c), calculated with scattering matrix truncated to the lowest $21$ diffraction orders ($-10\leqslant n \leqslant 10$).
In all three cases, the first ($n=1$) and minus first ($n=-1$) diffraction-order light lines are evident as dips in $1-T_0$ that move apart from each other almost linearly with $Q$, as $Q$ increases starting from zero.  When the contribution of the interband conductivity $\sigma_\text{inter}$ is removed from the total conductivity $\sigma$, these dips are clearly seen in \Fig{fig:enhance}(b) in addition to a step-like feature also moving linearly with $Q$ along the minus-first diffraction-order light line. 
This step-like feature, also present in some form in the graphene transmission [Figs.\,\ref{fig:enhance}(a) and (c)], is caused by opening the minus-first diffraction channel with increasing frequency in accordance with \Eq{open_channel_condition}. Such features are common to the optical spectra of periodic systems~\cite{BolotovskiiJETP68,WojcikPRL21}. In fact, as frequency increases to the point that \Eq{open_channel_condition} is satisfied for $n=-1$, the transmission of the corresponding ($n=-1$) diffraction order suddenly increases from zero to a finite value [see \Fig{fig:t5}(a) in Appendix \ref{transmission_appendix}] that results in a corresponding abrupt reduction in $T_0$, which is observed in \Fig{fig:enhance}(b).


The most important features in the spectra of periodically modulated graphene [Figs.\,\ref{fig:enhance}(a) and (c)] are also highlighted by their comparison with the normal metal spectra [Fig.\,\ref{fig:enhance}(b)]. These are peaks in $1-T_0$ which appear also close to the folded light lines and are the manifestation of the TE SPP modes in periodically modulated graphene.
At the frequencies of these TE SPP peaks, the energy of the incoming TE beam is diverted back into a reflection channel, as demonstrated in Appendix \ref{transmission_appendix}, showing in the same and in a much wider frequency range the transmission in zeroth and other ($n=\pm1$) diffraction orders, as well as the absorption, see \Figs{fig:t5}{fig:multit5} for $\Delta=0$ (results for non-zero detuning $\Delta$ are also presented in Appendix \ref{transmission_appendix}).

In the selected frequency range  close to $\Omega=G_1$, one would expect to observe  in Figs.\,\ref{fig:enhance}(a) and (c) two TE SPP modes, separated by a frequency gap, in accordance with \Fig{fig:folded_disp}(b). This is a general property of planar photonic-crystal structures which is usually caused by guided-mode folding and their hybridization due to the coupling between diffraction orders \cite{TikhodeevPRB02,fujita1998,yablonski2001}.  However, for $Q=0$, we see in \Fig{fig:enhance}(a) only one peak at around $\Omega-2\approx -1.76\times 10^{-4}$. The second peak is missing for $Q=0$ and appears only at nonzero values of $Q$. This is, however, in a full agreement with our analysis of the transmission, presented in \Sec{sec:approximation} in terms of the bright and dark TE SPP modes. The real part of their frequencies, $\Omega_b$ and $\Omega_d$, are also shown in \Fig{fig:enhance}(a) by vertical dash-dotted lines, demonstrating a good agreement with the peak positions. The linewidth of the dark mode is increasing with $Q$, and both modes are deviating from their positions at $Q=0$. We evaluate the frequency gap between the two TE SPP modes to be approximately $1.47\times 10^{-4}$. Note that a similar behaviour has been observed in transmission spectra of a distributed feedback microcavity structure \cite{fujita1998,yablonski2001} in which bright and dark modes originate from the interaction of the folded guided-mode dispersion lines.

\begin{figure}[!t]
\centering
\includegraphics[]{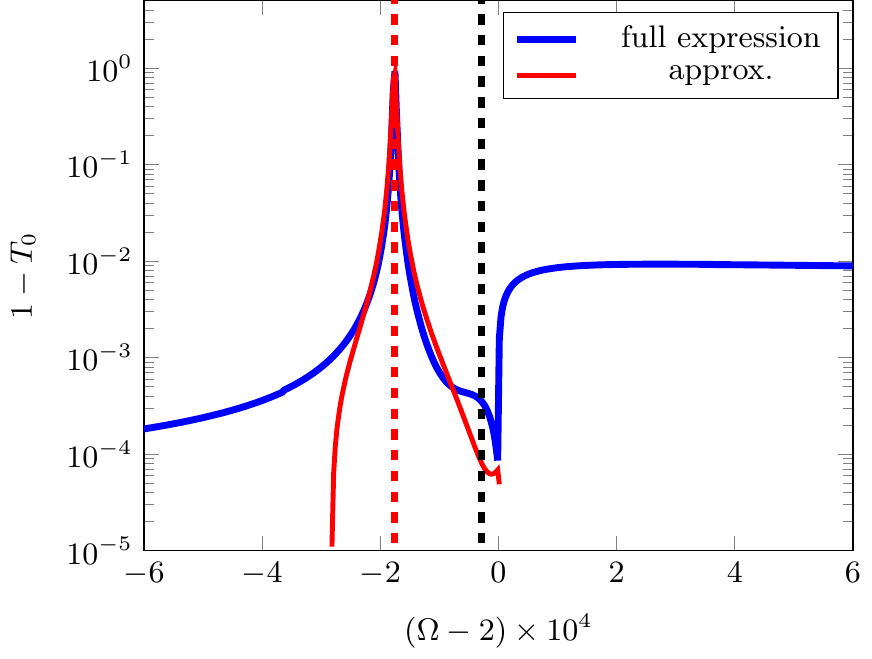}
\caption{Approximation of transmission (red solid line) using \Eq{tnotapprox} showing consistency with the TE peak for $Q=0$ and $\Delta=0$. Red and black dashed lines indicate the real part of the frequencies of the bright and dark TE SPP modes, obtained using \Eq{brightzero} and \Eq{darkzero}, respectively. The frequency gap between the dashed lines is approximately $1.47\times 10^{-4}$. }
\label{gratingtematch}
\end{figure}

A further analysis of the transmission spectrum at $Q=0$ is presented in \Fig{gratingtematch} which shows $1-T_0$ calculated using the approximation \Eq{tnotapprox} (red line) and the full scattering matrix truncated up to $21$ diffraction orders (blue line), demonstrating a good agreement between the two near $\Omega\approx\Omega_b$ (vertical red dashed line).  Away from $\Omega_b$, the approximate transmission \Eq{tnotapprox}, based on the first-order Taylor expansion, naturally deviates from the exact result.
This analysis provides a proof that the observed peaks in the transmission are in fact a manifestation of the TE SPP modes predicted in a homogeneous graphene layer~\cite{Mikhailov2007,Ahmad2021}. However, in homogeneous systems, these modes spectrally lie below the light line and thus cannot be excited resonantly. By periodically modulating the homogeneous graphene layer, the TE SPP modes can be seen in optical spectra.

\begin{figure}[]
\centering
\includegraphics[width=\columnwidth]{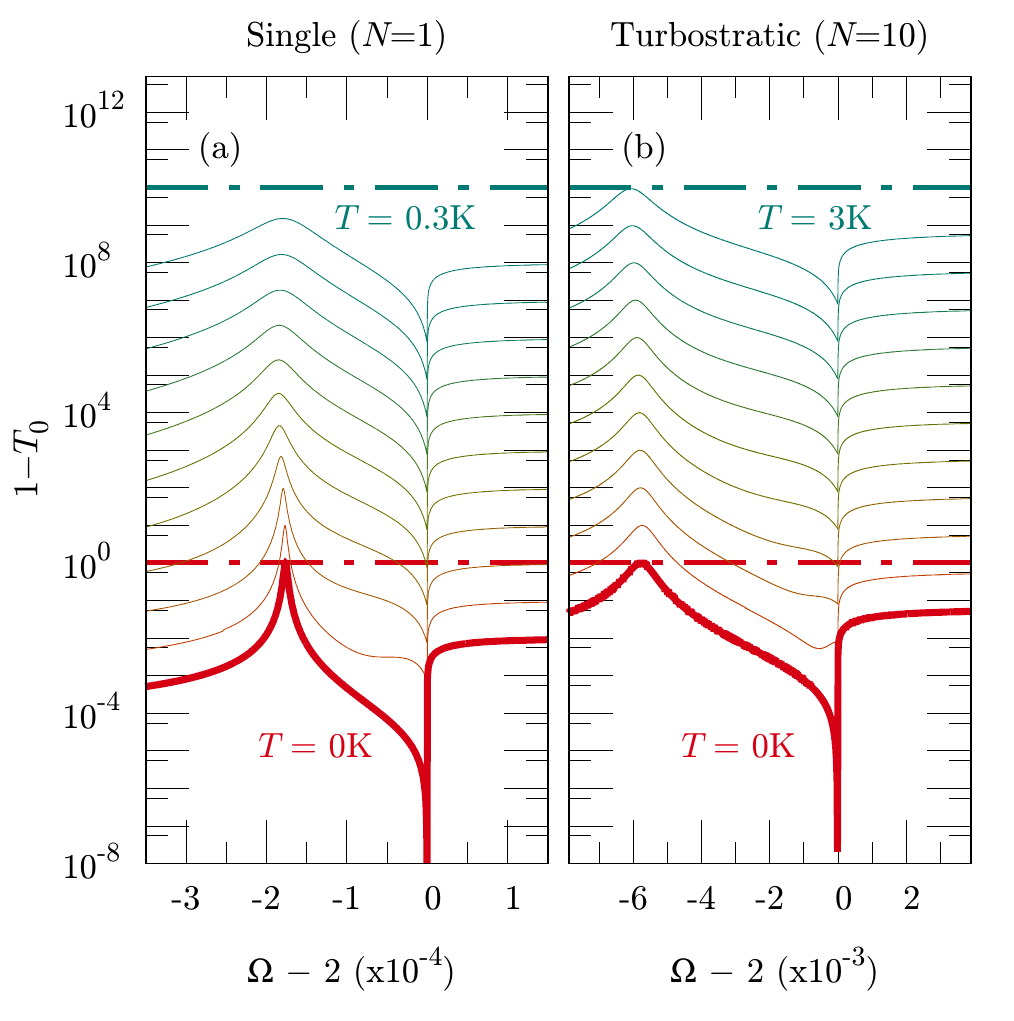}
\caption{Transmission spectra ($1-T_0$) for zeroth  diffraction order when a TE polarised light is normally incident ($Q=0$) on grating with $b/d=0.3$, for $\Delta=0$, shown from zero temperature (red) to finite temperature (green), for (a) single-layer graphene, and (b) turbostratically stacked graphene. Red and green dashed dotted lines indicate $T_0=0$ for zero temperature and $T=0.3K$ (single) or $T=3K$ (turbostratic), respectively. }
\label{fig:enhancement_T}
\end{figure}

Using turbostratic graphene grating as shown in \Fig{fig:enhance}(c), TE mode peaks coming from interband conductivity can be further pronounced. Similar to the single-layer case, the main peak  reaches its maximum value of 1. However, the width of the peaks due to TE SPPs and their positions in frequency are both affected by the use of turbostratic graphene. Note that the TE mode peak at $Q=0$ has been shifted further away from the light line, about a factor of $35$ more compared to its previous position for the single-layer case in \Fig{fig:enhance}(a). The peak is also wider in frequency making TE SPP mode easier to detect. The peak frequency does not change much for $Q$ close to zero, but eventually starts to shift to lower frequencies as $Q$ is increased (this can be seen in \Fig{fig:multit5}(c) of in Appendix \ref{transmission_appendix}). The dark TE SPP mode peak only appears for $Q\neq 0$ as seen in \Fig{fig:enhance}(c), similar to the single-layer case. However, it demonstrates a fine structure, consisting of multiple peaks, which are due to a stronger coupling between the TE SPP modes and higher diffraction orders more strongly contributing to the scattering matrix.

The effect of change in temperature on the transmission spectrum is presented in \Fig{fig:enhancement_T} for $Q=0$. Temperature appears in $\scatmat$ through the conductivity of graphene, where the increase in temperature has the smearing effect on the interband dip in the imaginary part of graphene conductivity~\cite{Ahmad2021}. Figure \ref{fig:enhancement_T} shows that the TE mode peak is reaching its maximum height of $1$ at zero temperature in both the single-layer and turbostratic case. For the single-layer case, increasing temperature smears the peak out, with the peak maximum reduced. For the turbostratic case, the peak is still reaching the maximum of $1$ at $3K$ with minimal smearing. This insensitivity of transmission to temperature suggests further in favour of turbostratic over single-layer graphene for applications using TE SPPs that aim to be robust against temperature change.

\section{Conclusions}
\label{Sec:summary}
We have shown that the complex-frequency transverse-electric surface plasmon polariton (TE SPP) modes in a graphene grating can be excited by external electro-magnetic waves with close-to-normal incidence. Specifically, a significant reduction in the zeroth diffraction-order transmission is observed near the interband transition frequency of graphene.
This reduction can be explained as a reflection of the energy of the incoming beam due to the resonant coupling to the TE SPP modes.
In addition, we have shown that the frequency and the in-plane wave number of TE SPP modes seen in transmission can be tuned by controlling the graphene chemical potential and grating period. Specifically, we have demonstrated that for zero detuning, that is the grating period is chosen so that the TE SPP mode energy matches the interband threshold of the graphene conductivity, the features in the optical spectra due to the TE SPPs are significantly enhanced.
Furthermore, by using turbostratic graphene layers with negligible interlayer interactions, we can enhance these features further, making them experimentally more accessible than in the single-layer graphene.

\section*{Acknowledgments}
Z.A acknowledgments financial support of EPSRC under the DTP scheme. S.S.O. acknowledges the support of S\^er Cymru II Rising Star Fellowship [80762-CU 145 (East)], which is partly funded by the European Regional Development Fund (ERDF) via the Welsh Government.

\appendix

\section{Scattering-matrix approach} \label{appendix_scat_mat}
\subsection{Solving Maxwell's equations for an infinitely thin periodic layer}
\newcommand{\Hz}{H_z}
\newcommand{\Hx}{H_x}
\newcommand{\Ey}{E_y}

We start from Maxwell's equations having the form
\begin{align}
    \nabla\times\mathbf{E}&=-\partial_t\mathbf{H}\,,\\
    \nabla\times\mathbf{H}&=\partial_t\mathbf{D}\,,
\end{align}
where we assume non-magnetic materials, i.e. the permeability $\mu=1$ everywhere in space, and use the units in which the speed of light $c=1$. 
Assuming a harmonic dependence on time of the electro-magnetic field in the form of $e^{-i\omega t}$, we obtain
\begin{align}
    \nabla\times\mathbf{E}&=i\omega\mathbf{H}\,,\\
    \nabla\times\mathbf{H}&=-i\omega\varepsilon(\omega)\mathbf{E}\,.
\end{align}
For the TE polarization, the magnetic and electric fields can be expressed in terms of their Cartesian components as:
\begin{align}
    \mathbf{H}=\begin{pmatrix}H_x\\0\\H_z\end{pmatrix}\,, \quad \mathbf{E}=\begin{pmatrix}0\\E_y\\0\end{pmatrix}\,.
\end{align}
Maxwell's equations then take the form
\begin{align}
    \begin{pmatrix}-\partial_z\Ey\\0\\\partial_x\Ey\end{pmatrix}&=+i\omega\begin{pmatrix}H_x\\0\\H_z\end{pmatrix}\,,\\
    \begin{pmatrix}\partial_y H_z\\\partial_z\Hx-\partial_x\Hz\\-\partial_y\Hx\end{pmatrix}&=-i\omega\varepsilon\begin{pmatrix}0\\E_y\\0\end{pmatrix}\,.
\end{align}
These can be rewritten as
\begin{align}
    -\partial_z\Ey&=i\omega\Hx\label{tenontriv1}\,,\\
    \partial_x\Ey&=i\omega\Hz\label{tenontriv2}\,,\\
    \partial_z\Hx-\partial_x\Hz&=-i\omega\varepsilon\Ey\label{tenontriv3}.
\end{align}

As defined in \Eq{struct_eps}, the permittivity in the whole space is given by $\varepsilon(x,z;\omega)= 1+\chi(\omega)\Lambda(x)\delta(z)$, where $\chi(\omega)$ is the 2D susceptibility of a thin (graphene) layer and $\Lambda(x)$ is a periodic function describing its spatial modulation, which can be represented by its Fourier series as
\begin{align}
    \Lambda(x)=\sum_nV_ne^{ig_nx}\,.
\label{appa12}
\end{align}

Integrating \Eqs{tenontriv1}{tenontriv3} over $z$ across the point $z=0$, we obtain
 \begin{align}
     -\evalat{\Ey}{z=0^+}+\evalat{\Ey}{z=0^-}&=0\label{te1}\,,\\
     \evalat{\Hx}{z=0^+}-\evalat{\Hx}{z=0^-}&=-i\omega\chi(\omega)\Lambda(x)\evalat{\Ey}{z=0}
     \label{te2}\,,
 \end{align}
where $0^{+(-)}$ is a positive (negative) infinitesimal. For the TE polarization, it is easier to eliminate $\Hx$ and write equations in terms of the $\Ey$ component having the following general form
\begin{align}
    E_y(x,z)=\sum_n e^{i(q+g_n)x}\times
    \begin{cases}
      C_ne^{ik_nz}+D_ne^{-ik_nz} & z> 0\,, \\
      A_ne^{ik_nz}+B_ne^{-ik_nz} & z< 0\,. \\
   \end{cases} \label{appa11}
\end{align}
in which $g_n$ and $k_n$ are defined, respectively, by \Eqs{eq_qn}{eq_kn}, and $q$ is the wave number along $x$.

Substituting \Eq{tenontriv1} into \Eq{te2}, we obtain
\begin{align}
    \evalat{\partial_z\Ey}{z=0^+}&-\evalat{\partial_z\Ey}{z=0^-}=-\omega^2\chi(\omega)\Lambda(x)\evalat{\Ey}{z=0},
\end{align}
which after using the series \Eqs{appa12}{appa11} becomes
\begin{align}
    \sum_n &ik_n(C_n-D_n-A_n+B_n)e^{i(q+g_n)x}\nonumber\\
    &=-\omega^2\chi(\omega)\sum_{mn}e^{iqx}e^{ig_{m+n}x}V_m(A_n+B_n).\label{appa14}
\end{align}
By shifting the indices of summation  $n\xrightarrow{}n-m$ in the right-hand side of \Eq{appa14}, we find
\begin{align}
 \sum_n &ik_n(C_n-D_n-A_n+B_n)e^{i(q+g_n)x} \nonumber \\& =-\omega^2\chi(\omega)\sum_{m,n}e^{iqx}e^{ig_{n}x}V_m(A_{n-m}+B_{n-m})\,,\label{appa16}
\end{align}
and then equating the coefficients at $e^{i(q+g_n)x}$ in \Eq{appa16}, we obtain
\begin{align}
 ik_n&(C_n-D_n-A_n+B_n)\nonumber \\&=-\omega^2\chi(\omega)\sum_{m}V_m(A_{n-m}+B_{n-m}) \,.\label{appa17}
\end{align}
Finally, shifting the indices again in the summation in \Eq{appa17}, by substituting $m\xrightarrow{}-m+n$, we arrive at
\be
 ik_n(C_n\!-\!D_n\!-\!A_n\!+\!B_n)\!=\!-\omega^2\chi(\omega)\sum_{m}V_{n-m}(A_{m}\!+\!B_{m}) \label{tem1}\,.
\ee
In addition to this, we find from the continuity of the electric field, \Eq{te1}, that
\begin{align}
    C_n+D_n=A_n+B_n\label{tem2} \,.
\end{align}
An infinite set of simultaneous equations given by \Eqs{tem1}{tem2} presents a general solution of Maxwell's equation in TE polarization for the infinitely thin periodically modulated layer.

\begin{figure}[t]
\centering
\includegraphics[width=\columnwidth]{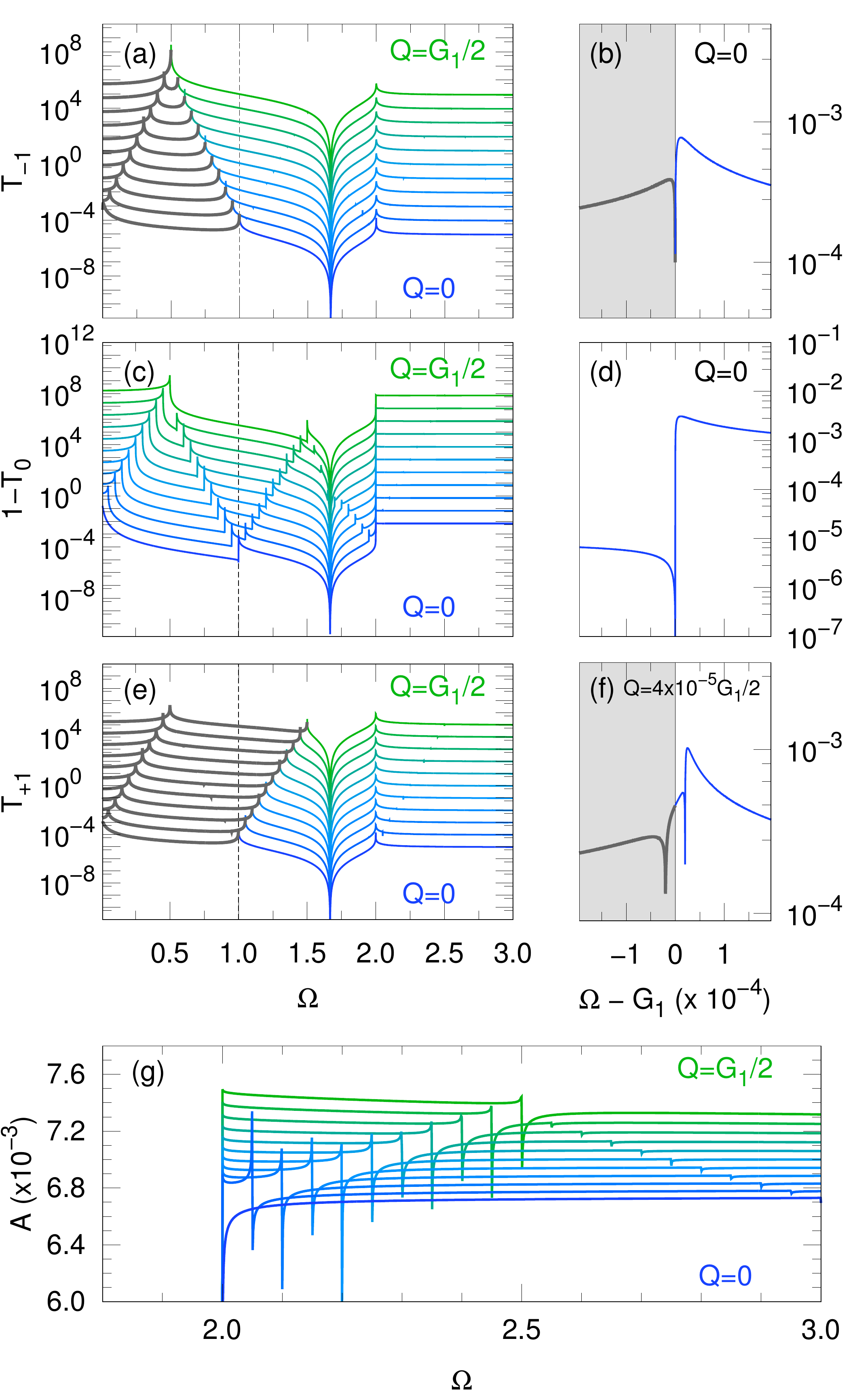}
\caption{Transmission spectra for single layer graphene ($N=1$), for the minus-first diffraction order (a),(b), zeroth order (c),(d), and first order (e),(f), for the period $d$ adjusted so that $\Delta=-1$ ($G_1=1$). Panels (b) and (d) zoom in the spectral features near $\Omega=G_1$ in (a) and (c), respectively, for $Q=0$. Panel (g) is absorption of open channels calculated using \Eq{absorp_form}, with waterfall increment of $+0.02$. $A(\Omega<2)=0$ for zero temperature. The gray solid lines correspond to the forbidden frequency range where \Eq{open_channel_condition} is not satisfied, and the curve has the meaning of the near-field amplitude.}
\label{fig:t1}
\end{figure}

\begin{figure}[]
\centering
\includegraphics[width=\columnwidth]{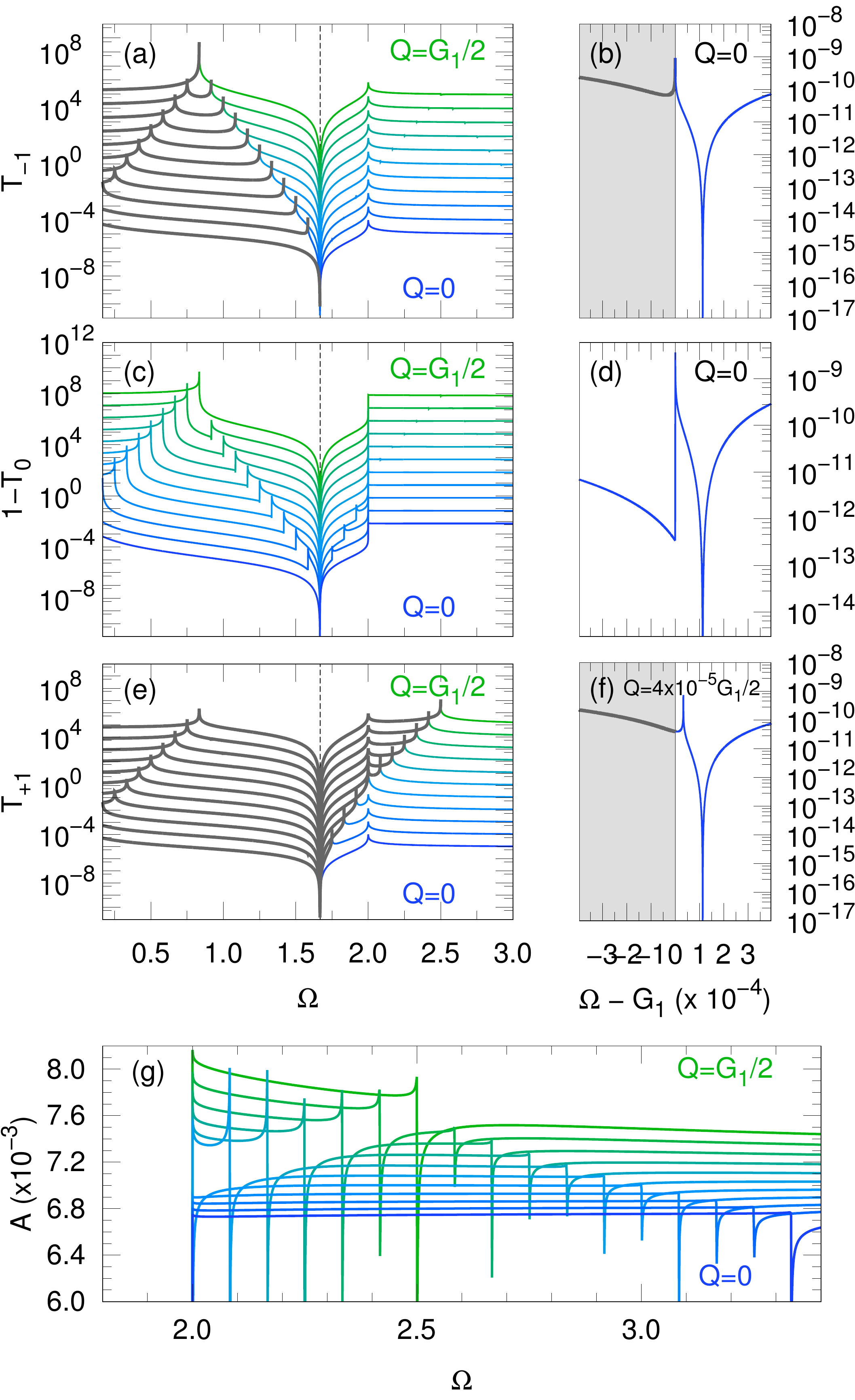}
\caption{As \Fig{fig:t1} but for $\Delta=-2+1.667$ ($G_1=1.667$), illustrating suppression of near-field and transmission amplitudes.}
\label{fig:t2}
\end{figure}

\begin{figure}[]
\centering
\includegraphics[width=\columnwidth]{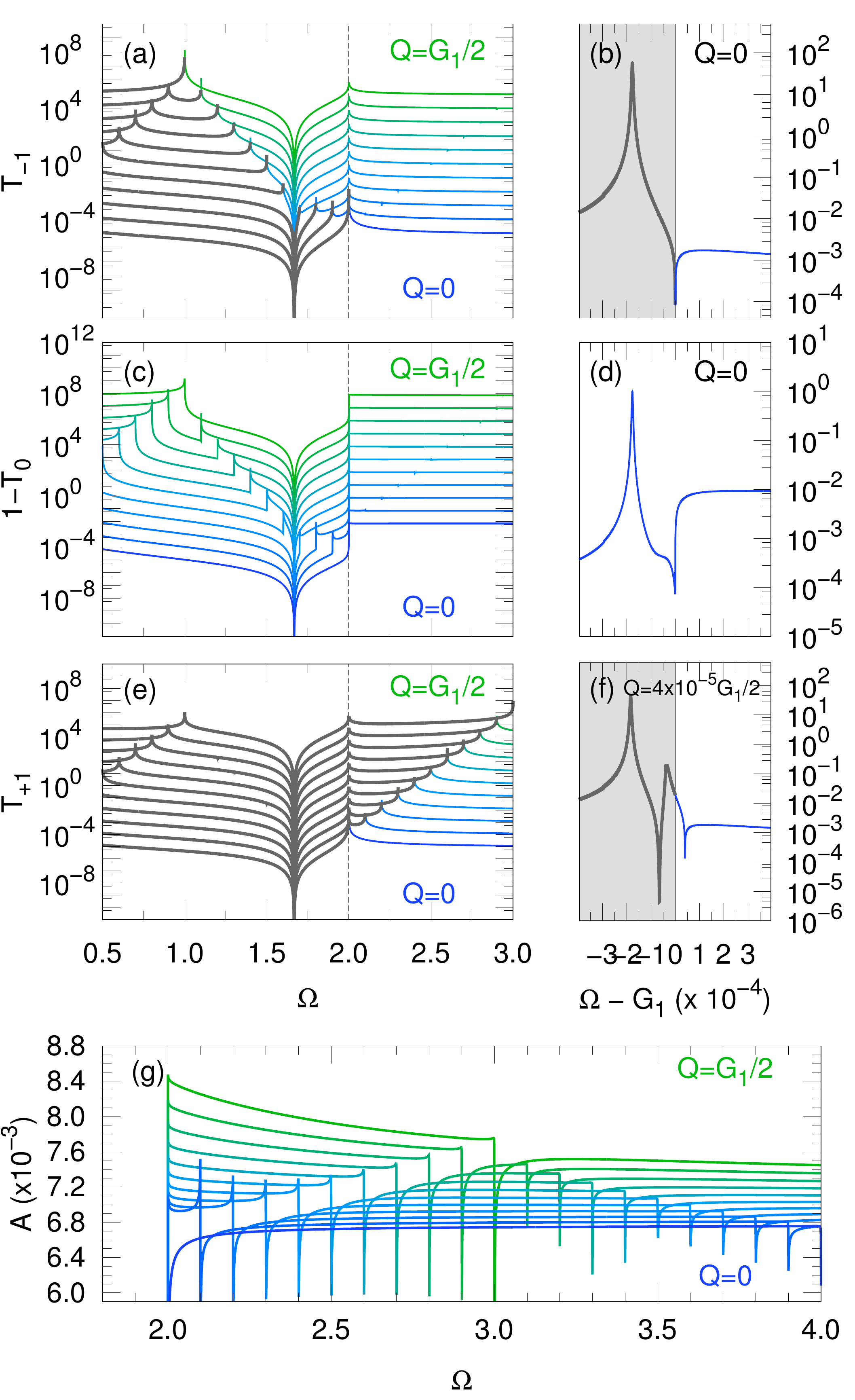}
\caption{As \Fig{fig:t1} but for $\Delta=0$ ($G_1=2$), illustrating the amplification of the near-field and transmission amplitudes.}
\label{fig:t5}
\end{figure}

\begin{figure}[]
\centering
\includegraphics[width=\columnwidth]{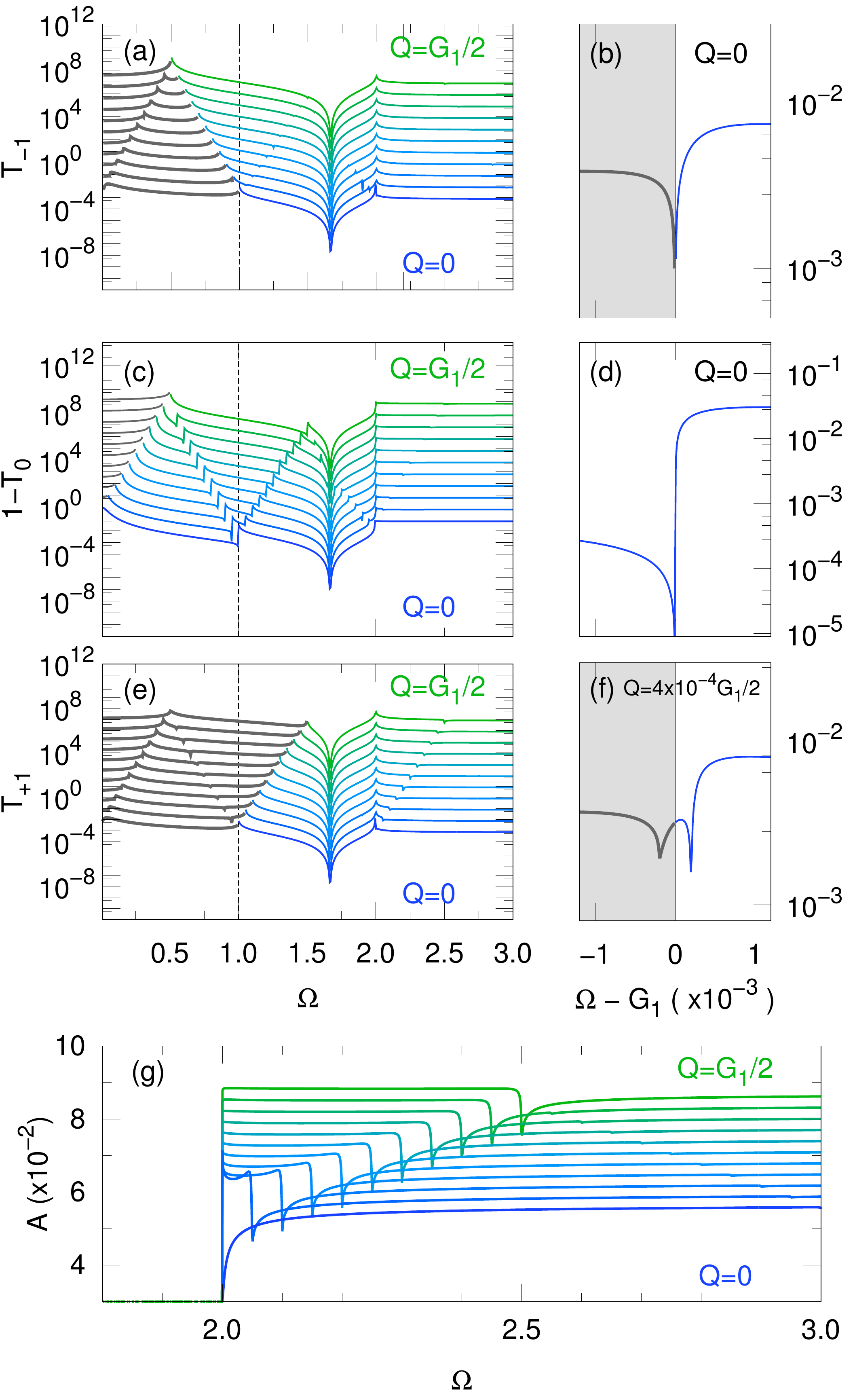}
\caption{Transmission spectra for turbostratic ($N=10$) graphene, for the minus first diffraction order (a),(b), zeroth order (c),(d), and first order (e),(f) for detuning $\Delta=-1$ ($G_1=1$). Panels (b) and (d) zoom in the spectral features near $\Omega=G_1$ in (a) and (c), respectively, for $Q=0$. Similarly, panel (f) is zoom in of panel (e) at $Q$ close to but not zero. Gray regions and gold curves indicate forbidden frequency regions. Panel (g) is absorption spectrum obtained using \Eq{absorp_form}, with waterfall increment of $+0.2$.}
\label{fig:multit1}
\end{figure}

\begin{figure}[]
\centering
\includegraphics[width=\columnwidth]{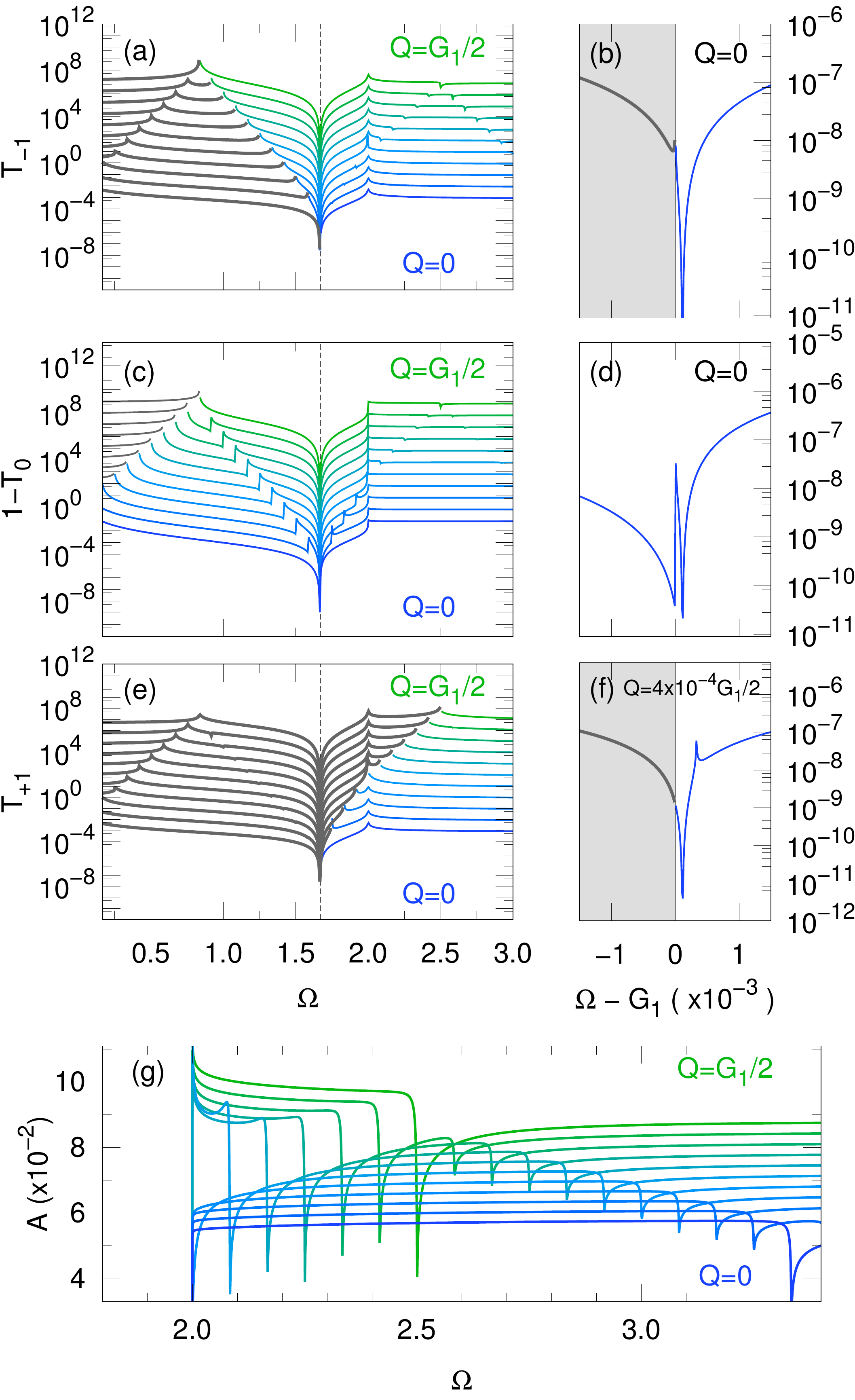}
\caption{As \Fig{fig:multit1} but for $\Delta=-2+1.667$ ($G_1=1.667$).}
\label{fig:multit2}
\end{figure}

\begin{figure}[]
\centering
\includegraphics[width=\columnwidth]{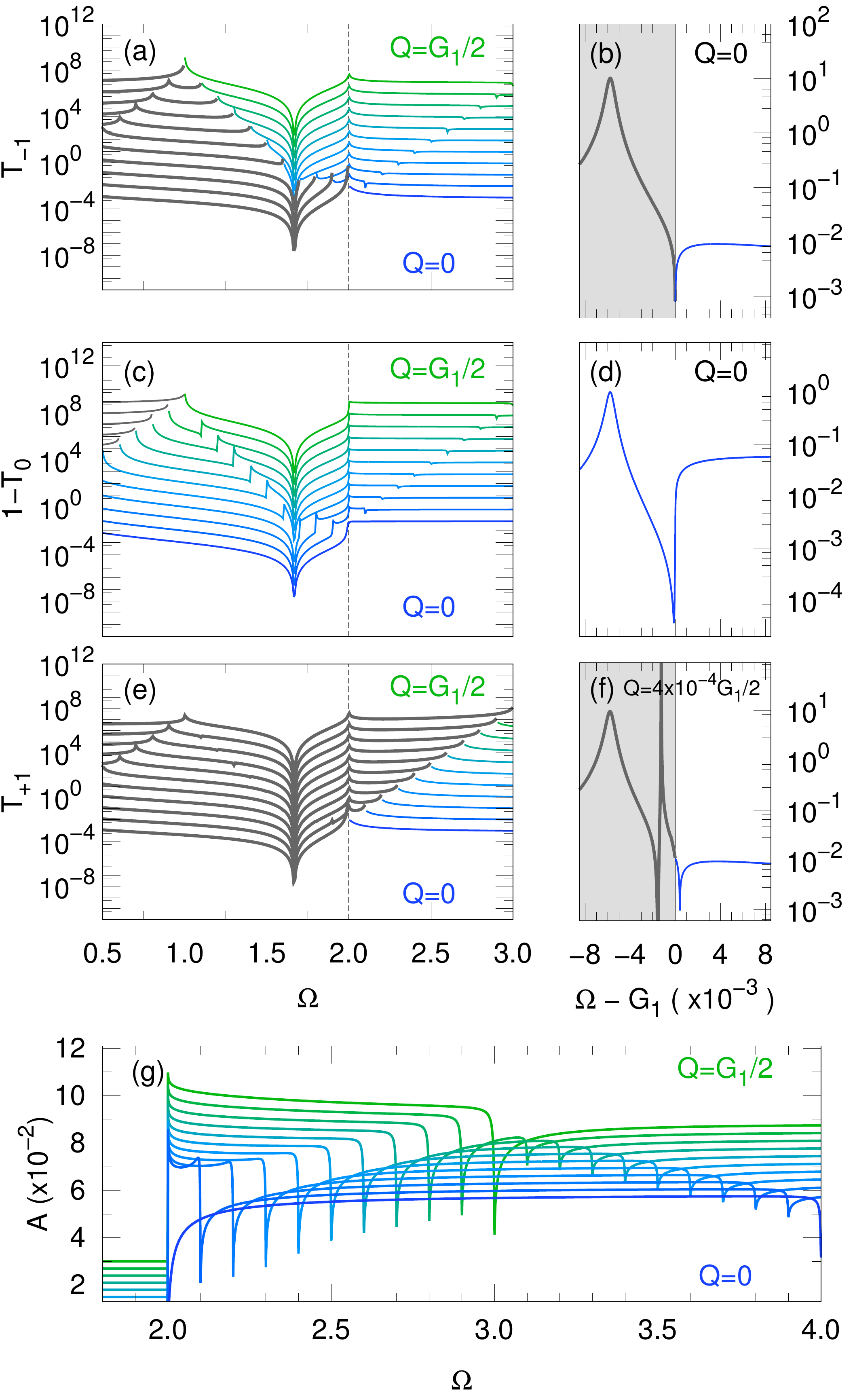}
\caption{As \Fig{fig:multit1} but for $\Delta=0$ ($G_1=2$).}
\label{fig:multit5}
\end{figure}

\begin{figure}[]
\centering
\includegraphics[width=\columnwidth]{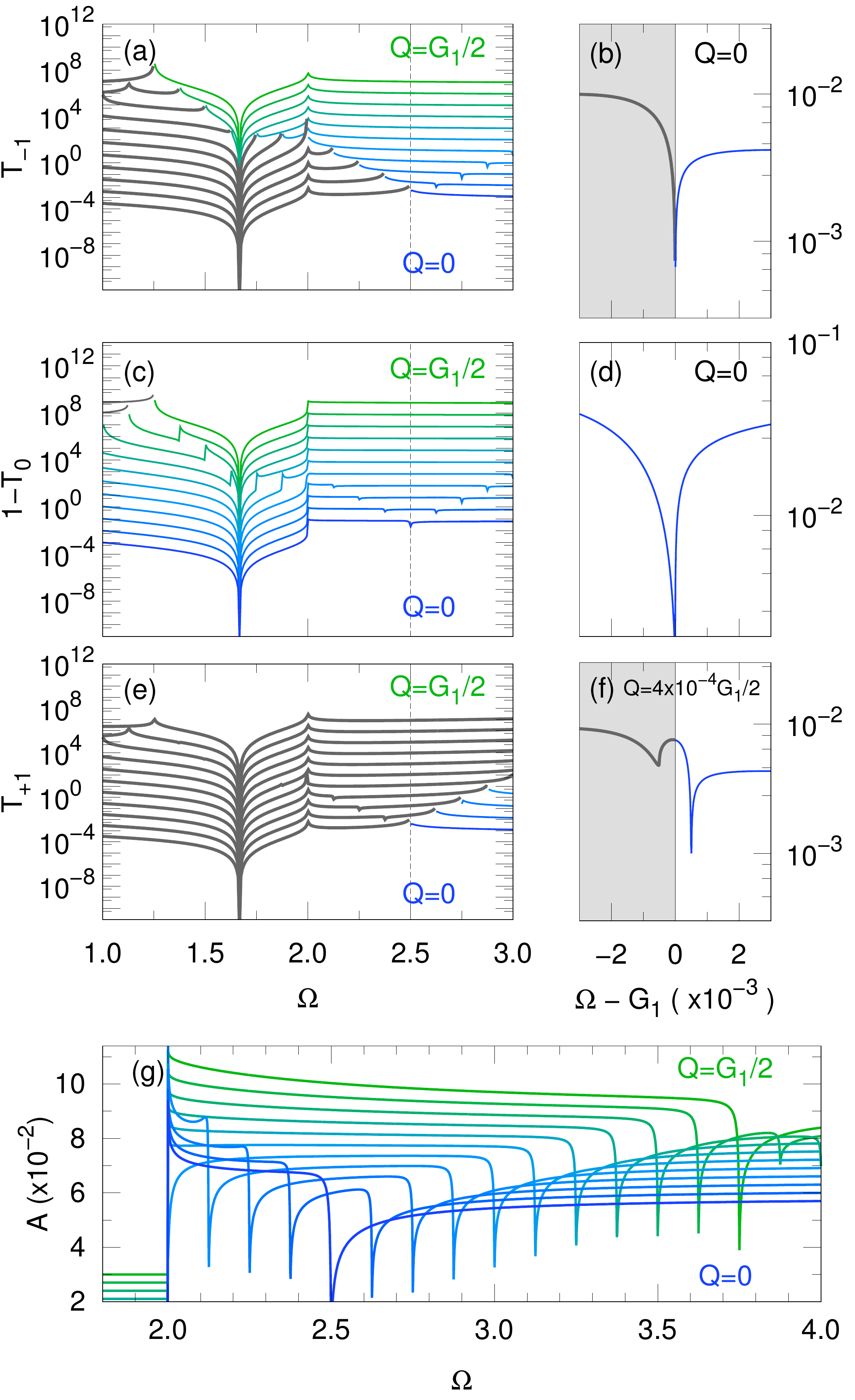}
\caption{As \Fig{fig:multit1} but for $\Delta=0.5$ ($G_1=2.5$).}
\label{fig:multit6}
\end{figure}

\subsection{Deriving the  scattering matrix}

To build a scattering matrix, let us rearrange \Eqs{tem1}{tem2} in the following way
\begin{align}
    C_n+B_n&+\sum_m\frac{\omega^2\chi(\omega)}{ik_n} V_{n-m}B_m\nonumber\\&=A_n+D_n-\sum_m\frac{\omega^2\chi(\omega)}{ik_n} V_{n-m}A_m\\
    C_n-B_n&=A_n-D_n
\end{align}
Thus, in terms of $2\times2$ matrices of diffraction-order blocks, the above equations can be written as
\begin{align}
    \begin{bmatrix} \I&\I+2\U\\\I&-\I\end{bmatrix}\begin{bmatrix} \mathbf{C}\\\mathbf{B} \end{bmatrix}=
    \begin{bmatrix} \I&\I-2\U\\-\I&\I \end{bmatrix}\begin{bmatrix} \mathbf{D}\\\mathbf{A} \end{bmatrix}\,,
\end{align}
where vectors $\mathbf{A}$, $\mathbf{B}$, $\mathbf{C}$, and $\mathbf{D}$ have components $A_n$, $B_n$, $C_n$, and $D_n$, respectively, matrix $\U$ has the matrix elements
\begin{align}
    U_{nm}=\frac{\omega^2\chi(\omega)}{2ik_n}V_{n-m}\,,
\end{align}
and $\I$ is the identity matrix.

The scattering matrix, relating the amplitudes of incoming and outgoing waves, is then given by
\begin{align}
    \scatmat=\begin{bmatrix} \I&\I+2\U\\\I&-\I\end{bmatrix}^{-1}
    \begin{bmatrix} \I&\I-2\U\\-\I&\I \end{bmatrix}.\label{sqmat_a25}
\end{align} 
To further simplify it, we can write the inverse of the square matrix in \Eq{sqmat_a25} in terms of its blocks:
\begin{align}
    \scatmat=\frac{1}{2}\begin{bmatrix} \Winv&\Winv(\I+2\U)\\\Winv&-\Winv\end{bmatrix}
    \begin{bmatrix} \I&\I-2\U\\-\I&\I \end{bmatrix},
\end{align} 
where $\W = \I+\U$. Finally, performing the matrix multiplication, we obtain
\begin{align}
    \scatmat=\begin{bmatrix} -\Winv\U&\Winv\\ \Winv&-\Winv\U\end{bmatrix} .\label{explicit_mat_result}
\end{align}

\subsection{TM polarization}
\label{appendix_TM}

We note that TM polarization can be treated in exactly the same manner as TE polarization solved above, leading to the same result for the scattering matrix, provided that matrix $\mathbf{U}$ is replaced with $\mathbf{U}^\text{TM}$ having the following matrix elements:
\begin{align}
    U_{nm}^\text{TM}= -\frac{1}{2}ik_n\chi(\omega) V_{n-m} \,.
\end{align}
Also, in TM polarization, the vectors $\mathbf{A}$, $\mathbf{B}$, $\mathbf{C}$, and $\mathbf{D}$ have the same meaning of the expansion coefficients in \Eq{appa11}, but for  $H_y(x,z)$ instead, which in this case is the only non-zero component of the magnetic field.

\section{Transmission and absorption spectra} 
\label{transmission_appendix}
Here we present the transmission spectra, calculated using \Eq{transmission_exp}, for grating with the filling factor $b/d=0.3$ for the first Brillouin zone $Q\in [0,G_1/2]$ and for different grating periods such  that $\Delta=-1$, $-2+1.667$, and 0. We also show the absorption defined as
\be
A(\Omega)= 1-\sum_{n\in\{\Omega>|Q+G_n|\}}\left[T_n(\Omega) + R_n(\Omega)\right]\,, \label{absorp_form}
\ee
where 
\be
    R_n(\Omega)= \left|\frac{K_n}{K_0}B_n^2\right|\label{reflection_exp}
\ee    
is the reflection coefficient of the $n$-th Bragg channel. Note that the summation in \Eq{absorp_form} is performed over open Bragg channels only, satisfying the condition \Eq{open_channel_condition}, as indicated in the formula.

\subsection{Single-layer case} \label{single_layer_case}

Figures~\ref{fig:t1}, \ref{fig:t2}, and \ref{fig:t5} show transmission and absorption spectra for single-layer ($N=1$) graphene grating for $\Delta=-1$, $-2+1.667$, and $0$, respectively, corresponding to $G_1=1$, $1.667$, and $2$.


The SPP modes manifest themselves as peaks in the transmission spectra in \Figsss{fig:t1}{fig:t2}{fig:t5} in panels (a), (c), and (e), and as dips in panel (g), all moving almost linearly with $Q$. 
The absorption $A$ in panel (g) is zero for $\Omega <2$ and is about $10^{-2}$ otherwise, which correlates with the interband absorption having a sharp frequency cut-off at $\Omega=2$ for zero temperature.
In the zeroth-order transmission $1-T_0$, the TE SPP mode peaks are lost due to the stronger interband absorption for frequencies above $\Omega>2$, instead appearing as dips in the absorption spectrum $A$, for such frequencies.
Since the mode features are sharp in frequency, we show zoom in of the TE mode transmission at $Q=0$ for minus-first [panels (b)] and zeroth [panels (d)] diffraction orders. First order transmission for small but finite $Q$ is shown in panels (f).

For $\Delta=-1$ (\Fig{fig:t1}), the TE mode dip (not peak) for $Q=0$ appears at frequency very close to $G_1$, see panel (b). When $Q$ is slightly increased [panel (f)], the TE mode dip moves away from $G_1$ increasing in frequency within the open channel region (not gray).

For $\Delta=-2+1.667$ (\Fig{fig:t2}), the TE mode peak can be also seen moving to higher frequencies for $Q$ slightly increasing from $Q=0$ [panel (b)] to $Q\neq0$ [panel (f)]. The peak width is very small, about $10^{-10}$ in this case due to the vanishing graphene conductivity near this frequency of $\Omega\approx 1.667$~\cite{Ahmad2021}. 

For $\Delta=0$ (\Fig{fig:t5}), the TE mode peak is clearly enhanced due to the dip in the imaginary part of the conductivity. The TE SPP peak in the coefficient of the evanescent near field (gray area) moves to lower frequencies for increasing $Q$ from panel (b) to (f), in agreement with the observations in \Fig{fig:enhance}(a).


\subsection{Turbostratic case}
Figures~\ref{fig:multit1}, \ref{fig:multit2}, \ref{fig:multit5}, and \ref{fig:multit6} show transmission and absorption spectra for turbostratic ($N=10$) graphene grating for $\Delta=-1$, $-2+1.667$, $0$, and $0.5$ respectively, corresponding to $G_1=1$, $1.667$, $2$ and $2.5$. Features related to the mode peaks and interband peaks are qualitatively the same here as in the single layer case in \Sec{single_layer_case} which can be seen by comparing panels (a), (c), (e), and (g) of all transmission spectra in this Appendix.

The factor of $N=10$ multiplying the conductivity increases the absorption by the same factor to the order of $10^{-1}$. The zoom in panels (b), (d), and (f) show effects of this factor in amplitude of peaks and the frequency of peaks and dips moving further away from $G_1$ by a factor of $10^2$, compared to the single-layer case. Importantly, the zoom in frequency near $\Omega=2$ in Figs.\,\ref{fig:multit5}(b), (d), and (f) shows evidence of enhancement in first and minus-first diffraction orders and a suppression to zero of the transmission ($1-T_0=1$) in the zeroth diffraction order.

\bibliography{export}

\end{document}